\documentclass{PoS}
\usepackage{amsmath}
\usepackage{graphicx}
\usepackage{txfonts}

\def\aap{A\&A}

\def\lum{\rm erg~s$^{-1}$}
\def\beq#1{\begin{equation}\label{#1}}
\def\eeq{\end{equation}}
\def\beqa#1{\begin{eqnarray}\label{#1}}
\def\eeqa{\end{eqnarray}}

\def\myfrac#1#2{\left(\frac{#1}{#2}\right)}

\def\comment#1{\relax}

\title{Low-luminosity stellar wind accretion onto neutron stars in HMXBs}

\ShortTitle{Low-luminosity stellar wind accretion onto neutron stars in HMXBs}

\author{\speaker{Konstantin Postnov}%
        \thanks{Supported by RSF grant 16-12-10519}\\
       Sternberg Astronomical Institute, Moscow M.V. Lomonosov State University, 
Universitetskij pr., 13,  Moscow 119234, Russia\\
       E-mail: \email{pk@sai.msu.ru}}

\author{Lida Oskinova\\
       Institute for Physics and Astronomy, University Potsdam, 
14476 Potsdam, Germany\\
        E-mail: \email{lida@astro.physik.uni-potsdam.de}}

\author{Jose Miguel Torrej\'on\\
       Instituto Universitario de F\'{\i}sica Aplicada a las Ciencias y las Tecnolog\'{\i}as, Universidad de Alicante, 03690 Alicante, Spain\\
        E-mail: \email{jmt@ua.es}}

\abstract{Features and applications of quasi-spherical settling accretion onto rotating magnetized neutron stars in high-mass X-ray binaries are discussed. The settling accretion occurs in wind-fed HMXBs when the plasma cooling time is longer than the  free-fall time from the gravitational capture radius, which can take place in low-luminosity HMXBs with $L_x\lesssim 4\times 10^{36}$~\lum. We briefly review the implications of the settling accretion, focusing on the SFXT phenomenon, which can be related to instability of the quasi-spherical convective shell above the neutron star magnetosphere due to magnetic reconnection from fast temporarily magnetized winds from OB-supergiant. If a young neutron star in a wind-fed HMXB is rapidly rotating, the propeller regime in a quasi-spherical hot shell occurs. We show that X-ray spectral and temporal properties of enigmatic $\gamma$~Cas Be-stars are consistent with failed settling accretion regime onto a propelling neutron star. The subsequent evolutionary stage of $\gamma$\,Cas and its 
analogs should be the X\,Per-type binaries comprising low-luminosity 
slowly rotating X-ray pulsars.}

\FullConference{11th INTEGRAL Conference Gamma-Ray Astrophysics in Multi-Wavelength Perspective,\\
		10-14 October 2016\\
		Amsterdam, The Netherlands}

\begin{document}

\section{Two regimes of wind accretion}

In high-mass X-ray binaries (HMXB), a compact stellar remnant (neutron star (NS) or 
black hole) accretes from stellar wind of a massive ($\sim 10-20 M_\odot$) early type optical companion 
\cite{2015A&ARv..23....2W}. Here we will consider accretion onto NS only. 
The stellar wind is captured by the orbiting NS at the characteristic capture (Bondi) radius $R_B=2GM_x/(v_{orb}^2+v_w^2)$, where $v_{orb}$ 
is the orbital NS velocity, $v_w$ is the stellar wind velocity at the NS distance from the optical star.
Typically, in HMXBs $v_w\sim 10^8\mathrm{cm~s}^{-1}\gg v_{orb}$. In the stellar wind, a bow shock is formed 
at about $R_B$. Downstream the shock, plasma falls towards the NS, and in the
case of rotating magnetized NS, is stopped by the NS magnetosphere at the characteristic distance $\sim R_A$, 
where $R_A$ is the magnetosphere (Alfv\'enic) radius determined by the pressure balance between the
falling plasma and the NS magnetic field. Plasma enters the NS magnetosphere via different (mostly Rayleigh-Taylor) instabilities, gets frozen into the magnetic field and is canalized towards the NS magnetic poles.
The accretion power is released near the NS surface with a total luminosity $L_x\sim 0.1 \dot Mc^2$, where
$\dot M$ is the accretion rate onto the NS.  
      
There can be two regimes of wind accretion onto magnetized NS in 
X-ray binaries: direct supersonic (Bondi) accretion, when the plasma captured  
from stellar wind cools down on time scales shorter than the free-fall time from $R_B$, $t_{cool}\ll t_{ff}(R_B)$, 
and quasi-spherical subsonic (settling) accretion, when the plasma cooling time 
is longer than the time of matter fall from $R_B$ to $R_A$,  $t_{cool}\gg t_{ff}(R_B)$. The cooling of plasma is the
most efficient via Compton interaction with X-ray photons produced near the NS surface, 
therefore the two regimes occurs at high (direct Bondi supersonic accretion) and low (subsonic 
settling accretion) X-ray luminosities. Transition between these accretion regimes occurs at $L_x\simeq 4\times 10^{36}$~erg~s$^{-1}$ \cite{2012MNRAS.420..216S}. Observational appearances of the settling accretion
regime are discussed in \cite{2015ARep...59..645S}. Importantly, in the settling accretion regime
the plasma entry rate in the NS magnetosphere is determined by the plasma cooling time and 
is less than than in the direct Bondi-Hoyle-Littleton case by the factor $f(u)=(t_{cool}/t_{ff})^{1/3}$, 
which can be much smaller than unity.  
 
\section{A model of bright flares in SFXTs}

Supergiant Fast X-ray transients (SFXTs) are a subclass of transient hard X-ray HMXBs with NSs discovered by INTEGRAL
\cite{2003ATel..190....1S,2005A&A...444..221S,2005ATel..470....1N}, which have been actively studied in the last years. They are characterized by sporadic
powerful outbursts on top of a low-luminosity quiescent state (see \cite{2014MNRAS.439.3439P} for a summary of the SFXT properties). 

At softer X--rays (1-10 keV), 
SFXTs are caught at X--ray luminosities below $L_q\sim 10^{34}$~erg~s$^{-1}$ most of the time.
A luminosity as low as 10$^{32}$~erg~s$^{-1}$ (1-10 keV) has been
observed in some sources, leading to a very high observed dynamic range (ratio between luminosity during outburst and quiescence)
up to six orders of magnitude (e.g.IGR\,J17544--2619 \cite{2015A&A...576L...4R}).

Low X-ray luminosity at quiescence suggests that quasi-spherical settling accretion can be realized in SFXTs.
In the frame of this accretion regime a model \cite{2014MNRAS.442.2325S} was proposed explaining the origin of powerful flares in these sources. Indeed, in the settling accretion regime a hot quasi-spherical convective shell is formed above the NS magnetosphere. The mass accretion rate through the shell is determined by the plasma cooling time,
$\dot M_a\approx f(u)\dot M_{B}$ (see above), and under stationary external conditions (density and velocity of the stellar wind) the shell can be in a quasi-steady state. Note that here changes in the wind density (possible clumping) lead to moderate variations of the resulting mass accretion rate, $\dot M\sim \rho^{11/32}v_w^{-33/16}$ (here we assume the most effective Compton cooling), while for the high-luminosity Bondi accretion we expect a much stronger dependence, $\dot M_B\sim \rho v_w^{-3}$. In the model \cite{2014MNRAS.442.2325S},  a powerful SXFT flare is produced when the entire shell above the NS magnetosphere is destabilized and accretes onto the NS in the free-fall time scale. As the mass of the shell in the quiescent state is $\Delta M \simeq \dot M_B t_{ff}\approx L_{q}/f(u) t_{ff}$, there must be a correlation between the energy emitted in a bright flare and the mean quiescent luminosity, $\Delta M \sim L_q^{7/9}$, which is indeed observed for bright SFXT flares \cite{2014MNRAS.442.2325S}.

The reason for the instability of the shell can be magnetized stellar wind of the OB-companion. Here two factors are important: (1) that a magnetized wind blob may have smaller velocity thus increasing the mass accretion rate
through the shell and (2), the most importantly, that the tangential magnetic field carried by the wind blob 
can initiate magnetic reconnection near the magnetospheric boundary thus opening the magnetic field lines and enabling free plasma entry. For efficient reconnection to occur, the time of plasma entry in magnetosphere (i.e., the characteristic instability time $t_{inst}\sim f(u)t_{ff}(R_A)$) should be comparable to or longer than the magnetic reconnection time, which is $t_{rec}=\epsilon_r t_A\simeq \epsilon_r t_{ff}(R_A)$  (here $t_A\sim t_{ff}(R_A)$ is the 
Alfv\'enic time at the magnetosphere boundary, $\epsilon_r<1$ is the magnetic reconnection efficiency). Therefore, the condition for reconnection can be written as $f(u)<\epsilon_r\sim 0.1$, which can naturally be met in low-luminosity HMXBs. Note that in high-luminosity HMXBs the magnetized stellar wind could not largely disturb the NS magnetosphere, as the time of plasma entry 
in the magnetosphere in this case can be shorter than the reconnection time.

\section{On the difference between persistent HMXBs and SFXTs}

Several mechanisms have been proposed to explain the SFXT phenomenon: accretion of dense stellar wind clumps \cite{2005A&A...441L...1I,2006ESASP.604..165N}, 
centrifugal inhibition of accretion at the magnetosphere (the propeller mechanism) \cite{2007AstL...33..149G,2008ApJ...683.1031B}, and wind accretion instability at the settling regime discussed in the previous Section \cite{2014MNRAS.442.2325S}. Optical companions of SFXTs and persistent HMXBs are very similar (e.g. \cite{2006ESASP.604..165N}, and the different behaviour of NSs in these binaries are apparently related to the stellar wind properties. Recently, \cite{2016A&A...591A..26G}  performed a comparative study of optical companions in two prototypical sources of these classes, IGR J17544-2619 (O9I) and Vela X-1 (B0.5Iae).  It was found that the wind termination velocity in IGR J17544-2619 ($v_\infty\simeq 1500$~km~s$^{-1}$) is two times as high as in Vela X-1 ($\simeq 700$~km~s$^{-1}$). The high wind velocity strongly decreases the captured mass accretion rate, since in the radiation-driven winds from early type supergiants $\dot M_B\sim \dot M_{OB}/v_\infty^4\sim L_{OB}/v_\infty^5$, where $L_{OB}$ is the star's luminosity. The authors conclude that if the NS in IGR J17544-2619 is indeed rapidly rotating (tens of seconds), the propeller mechanism is most likely. However, in other SFXTs with slowly rotating NSs (e.g., in IGR J16418-4532 with $P^*=1212$~s and IGR J16465-4507 with  $P^*=228$~s the applicability of the propeller mechanism may be questionable.

We suggest that the possible key difference between the SFXTs and persistent HMXBs is related to fast magnetized stellar winds of OB-supergiants in SFXTs. Magnetic fields are directly observed in about $10\%$ of OB-stars, and the magnetic field can significantly affect the stellar wind properties \cite{2011A&A...534A.140C}. 
Recently, the cumulative
distributions of the waiting-time (time interval between two consecutive X-ray flares), and
the duration of the hard X-ray activity (duration of the brightest phase of an SFXT outburst),
as observed by INTEGRAL/IBIS in the energy band 17-50 keV detected from
three SFXTs showing the most extreme transient behaviour
(XTE J1739-302, IGRJ17544-2619, SAXJ1818.6-1703) were analyzed \cite{2016MNRAS.457.3693S}.
A correlation between
the total energy emitted during SFXT outbursts and the time interval covered by the outbursts
(defined as the elapsed time between the first and the last flare belonging to the same
outburst as observed by INTEGRAL) was found. It was suggested that temporal properties of flares and outbursts
of the sources, which share common properties regardless different orbital parameters,
can be interpreted in the model of magnetized stellar winds with fractal structure from the
OB-supergiant stars.

To further test this hypothesis, it would be important to search for traces of magnetized stellar winds in spectra of the optical components of SFXTs.

\section{Non-accreting NSs in quasi-spherical shells: case of $\gamma$ Cas stars}

\textbf{Features  of propelling NSs at quasi-spherical settling accretion stage}.
As mentioned above, at low accretion rates in HMXBs with rapidly rotating NSs the propeller mechanism can operate. A young rapidly rotating NS formed after supernova explosion in a HMXB should spin-down at the propeller stage before accretion can be possible.  
For the conventional disk accretion, the duration of the propeller stage 
is determined by the NS spin period $P^*$, size of the NS magnetosphere $R_A$ and the NS magnetic field (dipole magnetic moment $\mu$), $\Delta t_P\sim \mu^{-2} R_A^3 (P^*)^{-1}$.  For typical parameters, the propeller stage lasts a hundred thousand years. 
In the settling accretion the NS spin-sown is mediated by the hot convective shell around the magnetosphere,
the propeller stage duration $\Delta t_P$ strongly depends on the wind velocity and binary orbital period,  
$\Delta t_P\sim v_w^7 P_{orb}^2L_x^{-1}$ and can be very long 
\cite{2017MNRAS.465L.119P}, of order of million years, which is comparable to the life time of the optical companion. Therefore, a sizable fraction of young rapidly rotating NSs in  HMXBs can be at the propeller stage. This is especially relevant to low-kick NSs originated from e-capture supernovae (ECSN)\cite{2004ApJ...612.1044P,2010NewAR..54..140V}. Indeed, in this case the NS remains in the orbital plane of the binary system, and if the optical companion is the Be-star that acquired fast rotation during the preceding mass transfer stage, the NS can be immersed in the low-velocity dense equatorial Be-disk outflow. Then the quasi-spherical settling accretion onto NS can be realized.

Observational manifestations of the propeller mechanism in wind-fed sources with settling accretion were considered in \cite{2017MNRAS.465L.119P}. 
The basic properties of the hot magnetospheric shell 
supported by a propelling NS in circular orbit in a binary system around a Be-star 
predicts the following observables: 

\smallskip\noindent
i) the system emits optically thin multi-temperature thermal radiation with the 
characteristic temperatures at the shell base above $\sim 10$\,keV: 
\begin{eqnarray}
\label{T_A}
&T_{\rm A}=\frac{2}{5}\frac{GM_{\rm X}}{{\cal R}R_{\rm A}}\approx 
27[\mathrm{keV}](R/10^9[\mathrm{cm}])^{-1}\nonumber \\
&\approx 
36[\mathrm{keV}]\mu_{30}^{-8/15}L_{32}^{2/15}(M_{\rm X}/M_\odot)^{19/15}\,,
\end{eqnarray}
where $\mu_{30}=\mu/(10^{30}\mathrm{G\,cm}^3)$ is the NS dipole magnetic moment, $L_{32}=L_{\rm X}/(10^{32}\mathrm{erg\, s}^{-1})$ is the X-ray luminosity of the shell. 
The shell is characterized by high plasma densities $\sim 10^{13}$~cm$^{-3}$:
\begin{equation}
\label{rho_m}
\rho_{\rm A}\approx 4.4\times 
10^{-11}[\mathrm{g\,cm}^{-3}](L_{32}/\mu_{30})^{2/3}(M_{\rm X}/M_\odot)^{1/3}\,,
\end{equation}
 and emission measures $\sim 10^{55}$~cm$^{-3}$:
\begin{eqnarray}
\label{ME}
&{\rm EM}=\int_{R_{\rm A}}^{R_{\rm B}}n_{\rm e}^2(r)4\pi r^2dr=4\pi 
n_{\rm e,A}^2R_{\rm A}^3\ln\myfrac{R_{\rm B}}{R_{\rm A}}\nonumber \\
&\approx 3.7\times 10^{54}[\mathrm{cm}^{-3}]
\mu_{30}^{4/15}L_{32}^{14/15}(M_{\rm X}/M_\odot)^{-2/15}\ln\myfrac{R_{\rm 
B}}{R_{\rm A}}\,.
\end{eqnarray}

\smallskip\noindent
ii) the typical X-ray luminosity of the system  is $L_X\sim 10^{33}$\,erg\,s$^{-1}$;

\smallskip\noindent
iii) no X-ray pulsations are present and no significant X-ray outbursts are 
expected in the case of a coplanar circular orbit with the Be-disk;

\smallskip\noindent
iv) the hot shell is convective and turbulent, therefore the observed X-ray emission lines 
from the optically thin plasma should  be broadened up to $\sim 1000$\,km\,s$^{-1}$;

\smallskip\noindent
v) the typical size of the hot shell is $\sim R_B \lesssim R_\odot$;

\smallskip\noindent
vi) the cold material, such as the Be-disk in the vicinity of 
the hot shell, should give rise to fluorescent FeK-line; 

\smallskip\noindent
vii) the life-time of a NS in the propeller regime in binaries with long orbital periods 
can be  $\sim 10^6$\,yrs, hence such systems should be observable among faint X-ray binaries in the Galaxy. 
\vskip\baselineskip

\textbf{Enigmatic X-ray emission from Be-star $\gamma$~Cas}.

The optically brightest Be-star $\gamma$\,Cas (B0.5IVpe) is well seen in the 
night 
sky by a naked eye even in a big city. It is a well known binary system with an 
orbital period of $P_b\approx 204$~d that consists of an optical star with mass 
$M_o\approx 16M_\odot$ and an unseen companion with mass $M_{\rm X}\approx 
1 M_\odot$ \cite{2000A&A...364L..85H,2002PASP..114.1226M}. The binary orbital plane and the disc of the Be-star are 
coplanar. The low-mass companion of $\gamma$ Cas can be a 
NS in the propeller stage grazing the outer Be-star disc region. 
The lack of periodic pulsations, as well as properties of the hot dense thermal plasma deduced from X-ray observations ($kT_{\rm hot}\sim 20$~keV and  $n_{\rm e}\sim 
10^{13}$~cm$^{-3}$, $v_{turb}\sim 1000$~km~s$^{-1}$)
(see \cite{2010A&A...512A..22L} and below), which challenge all previously proposed scenarios
of X-ray emission from $\gamma$~Cas \cite{2015arXiv151206446S}, are naturally 
expected in a hot magnetospheric shell around the propelling NS.

The propeller model explains both the gross X-ray features of $\gamma$\,Cas and well matches the detailed observations of its X-ray variability.
A hot convective shell above the NS magnetosphere should 
display time variability in a wide range, that 
depend on the sound speed, $c_s$, which is of the order of the free-fall time, $t_{\rm ff}$. 
The shortest time-scale is
$t_{\rm min}\sim R_{\rm A}/c_{\rm s}\sim R_{\rm m}/t_{\rm ff}(R_{\rm m})\sim 
R_{\rm m}^{3/2}/\sqrt{2GM}\sim$ a few seconds, while the longest time scale is $t_{\rm max}\sim 
R_{\rm B}/t_{\rm ff}(R_{\rm B})\sim 10^6[\mathrm{s}] (v_w/100 \rm{km\,s}^{-1})^{-3}\sim$ a few 
days. These are typical time scales of X-ray variability seen in 
$\gamma$\,Cas \cite{2010A&A...512A..22L}. 
The single power-law spectrum $P(f)\sim 1/f$  over the wide 
frequency range from 0.1\,Hz to $10^{-4}$\,Hz, as derived for the 
X-ray time variability in $\gamma$\,Cas 
\cite{2010A&A...512A..22L}, is common in accreting X-ray binaries    
\cite{2009A&A...507.1211R} and is thought to 
arise in turbulized flows beyond the magnetospheric boundary.

\begin{figure}
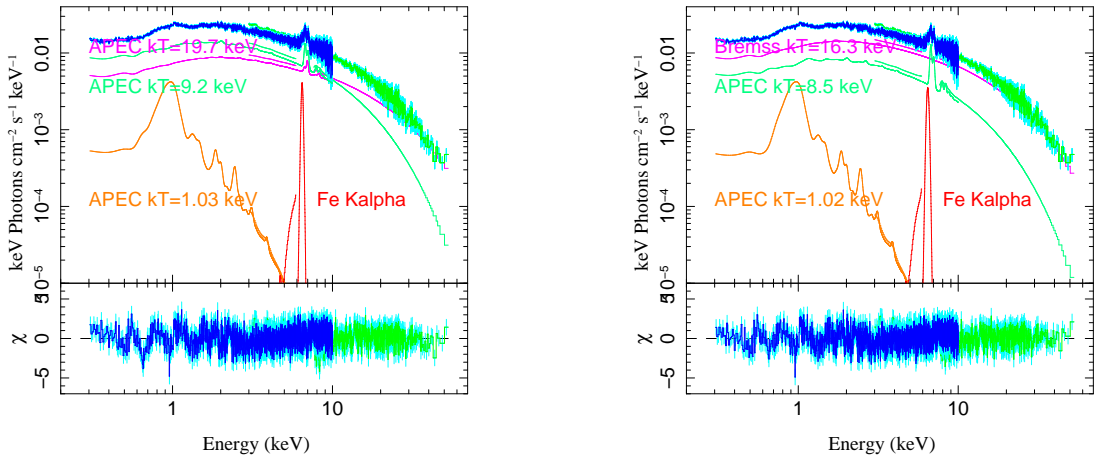

\includegraphics[width=0.45\textwidth]
{3bvapec_components_new.ps}
\hfill
\includegraphics[width=0.45\textwidth]
{bremss+2bvapec_components_new.ps}
\caption{Combined \textit{XMM-Newton} and NuSTAR X-ray spectra of $\gamma$~Cas fitted by
three APEC plasmas with under solar Fe abundance (Model 1, left panel) and 
thermal bremsstrahlung and two APEC components with solar Fe abundance 
(Model 2, right panel) and parameters listed in Table 1.}
\label{f:xray}
\end{figure}

\textbf{\textit{XMM-Newton} and NuSTAR observations of $\gamma$ Cas}.
We observed $\gamma$ Cas on 24-07-2014, simultaneously with {\it XMM-Newton} (ObsID
0743600101) and {\it
  NuSTAR} (ObsID 30001147002) space telescopes
(P.I. J.M. Torrej\'{o}n). The X-ray telescopes acquired data for 34 and 30.8 ks, respectively, providing for the first time 
a continuous spectral coverage from 0.3 to 60 keV. Unlike all previous X-ray 
observations, the low and high energy spectra are strictly simultaneous, overlap from 3 to
10 keV, with no gaps, and have S/N ratios at high energies which allow to
constrain the high energy spectral components, minimizing any possible
contribution from long term variability. A full analysis of
these data sets is under way and will be presented elsewhere. 
Here we concentrate mostly on the high energy continuum 
and the most prominent emission lines. In Fig. \ref{f:xray} we present the data
sets rebinned to have minimum S/N ratio per bin of 5. Besides the
continuum, three emission lines are clearly seen: Fe XXV He like at 6.68 keV, Fe XXVI H-like Ly
$\alpha$ at 6.97 keV. These transitions arise in a very hot thermal
plasma (see below). A third Fe line at 6.39 keV arises from cold near
neutral Fe fluorescence. The Fe K$\alpha$ transition is not included
in the thermal plasma models and is modelled with a Gaussian. 

In Table 1 we present the results of our analysis. 
We have essentially two ways of describing the whole spectrum: 
with three APEC plasmas at (19.7, 9.2 and ~1 keV) and Fe sub solar
abundance (~0.5 $Z_{\odot}$) -- Model 2 -- 
or one bremsstrahlung ($kT$=16.7 keV) and two APECs (8.5, ~1 keV) and
solar Fe abundance -- Model 1. 
The first solution corresponds, essentially, to what has been said before in previous studies. 
The sub solar abundance is driven by the fact that the 
continuum requires a high temperature and this over predicts 
the intensity of the Fe hot lines (Fe XXV and Fe XXVI). 
This Fe under abundance is neither predicted nor explained in any of
the competing explanations. 

The Model 1 solution, in turn, allows Fe solar abundance. The
bremmstrahlung component is the main contributor at all energies and
becomes dominant above 20 keV. The volume emission
measures, deduced from the normalization constants $C_{i}$ in Table 1 are
$EM\sim 4.7\times10^{55}$~cm$^{-3}$ for the Bremsstrahlung and 
$2.2 \times 10^{55}$ cm$^{-3}$ for the hot APEC component, respectively. 
For the Model 2, the EMs seem to be more similar, $2.6\times 10^{55}$ and $3.7\times 10^{55}$ cm$^{-3}$ for the APEC 1 and 2, respectively.

In terms of reduced $\chi^{2}$ Model 1 is better than Model 2 although the difference is not large (1.237 vs 1.284). 
The unabsorbed X-ray luminosity observed is $L_x = 1.5 \times 10^{33}$
erg~s$^{-1}$. Both models provide acceptable descriptions of the
observational data. Except for the Fe line under abundance for the first solution, 
the rest
of the parameters can be perfectly explained within the failed
quasi-spherical accretion paradigm presented in \cite{2017MNRAS.465L.119P}.

\begin{table}
\begin{tabular}{lll}
\hline
 Parameter & Model 1 & Model 2\\
\hline
      & &\\
N$_{H}^{is}$ ($10^{22}$ cm$^{-2}$) & 0.100$\pm 0.004$ &  0.104$^{+0.004}_{-0.005}$ \\
cf & 0.71$\pm 0.02$ &   0.72$\pm 0.03$  \\
N$_{H}^{disk}$ ($10^{22}$ cm$^{-2}$) & 1.12$^{+0.12}_{-0.10}$ &
1.21$\pm 0.17$ \\
      & &\\
C$_{Bremss}$ &  0.036$\pm 0.001$   &   ...\\
kT$_{Bremss}$ (keV) &  16.3$\pm 0.3$ & ...  \\
      & &\\
C$_{APEC 1}$ &  ...  &   0.065$^{+0.017}_{-0.008}$ \\
kT$_{APEC 1}$ (keV) &  ...  &  19.7$^{1.8}_{-1.6}$ \\
$Z_{Fe}$ ($Z_{\odot}$) & 1 & 0.48$\pm 0.02$ \\
      & &\\
C$_{APEC 2}$ &   0.053$\pm 0.006$   &  0.091$^{+0.009}_{-0.017}$  \\
kT$_{APEC 2}$ (keV) &  8.5$\pm 0.5$ &  9.2$^{+0.5}_{-0.4}$ \\
      & &\\
C$_{APEC 3}$ &  0.0026$\pm 0.0002 $   &   0.0026$\pm 0.0003$ \\
kT$_{APEC 3}$ (keV) &  1.019$\pm 0.016$ &   1.028$\pm 0.017$\\
      & &\\
$\lambda_{Fe K\alpha}$ (\AA) &  1.922$^{+0.005}_{-0.004}$  &   1.922$^{+0.005}_{-0.004}$\\
$I_{Fe K\alpha}$ (10$^{-5}$ ph s$^{-1}$ cm$^{-2}$) &  17$\pm 2$  &  15$\pm 2$ \\
$\sigma_{Fe K\alpha}$ (\AA) &  0.021$^{+0.007}_{-0.008}$    & 0.021$^{+0.007}_{-0.008}$   \\
      & &\\
$\chi_{r}^{2}$ (dof) &  1.237 (3129-14)    &   1.284 (3129-14) \\
\hline
\end{tabular}
\caption{Best parameters from the simultaneous fit to \textit{XMM-Newton} and NuSTAR
  data. All errors are quoted at the 90\% confidence level.}
\end{table}

\section{Conclusion} 

1) Accretion onto slowly rotating ($P^*\gtrsim 100$~s) magnetized NSs in wind-fed HMXBs can proceed in two different regimes -- direct Bondi-Hoyle-Littleton supersonic accretion, which is realized at high X-ray luminosities $L_x\gtrsim 4\times 10^{36}$~\lum, and settling subsonic accretion at lower luminosities. In the last case, a quasi-spherical convective shell is formed around the magnetosphere, which mediates the angular momentum transfer to/from the NS \cite{2012MNRAS.420..216S}. The settling accretion model can explain the low-states observed in accreting X-ray pulsars, the long-term spin-down in GX 1+4 and the existence of a very slowly rotating NSs without need of superstrong magnetic fields \cite{2015ARep...59..645S}.

2) The quasi-spherical shell formed at the settling accretion stage is quasi-stationary, but can be destabilized by magnetic reconnection if accretion occurs from magnetized stellar wind of the early type supergiant companion. This provides the possible explanation to SFXT bright flares \cite{2014MNRAS.442.2325S}. If the wind from the OB-supergiant in an HMXB with slowly rotating NS is fast and temporarily magnetized, the SFXT phenomenon can be observed, while 
if the wind is slow, a brighter persistent HMXB like Vela X-1 will appear (here the wind magnetization is unimportant since the time for magnetic reconnection is longer than the time it takes for plasma to enter the NS magnetosphere). 

3) The spectral and timing properties of the hot multi-temperature plasma filling the quasi-spherical convective shell above NS magnetosphere, which is formed at the settling accretion stage, were shown to be consistent with X-ray spectral properties of enigmatic low-luminosity Be X-ray binary $\gamma$ Cas and its analogs \cite{2017MNRAS.465L.119P}. Therefore, $\gamma$ Cas systems can be propelling NSs inside quasi-spherical shells at subsonic settling accretion stage
from Be-disks. These NSs were formed from low-kick ECSN supernovae in Be-binaries. With time, the propelling NS in $\gamma$ Cas and its analogs should spin down and will become 
slowly rotating X-ray pulsars at the stage of quasi-spherical settling 
accretion with moderate X-ray luminosity like X Per \cite{2012MNRAS.423.1978L}

Given the significant time duration, the propelling NSs surrounded by hot quasi-spherical shells can 
be present among low-luminosity non-pulsating HMXBs. 
Their X-ray spectral properties as summarized above and their long orbital 
binary periods 
can be used to distinguish them from, for example, faint hard X-ray emission from 
magnetic cataclysmic variables (e.g. \cite{2016ApJ...825..132H}).

\vskip\baselineskip
\textbf{Acknowledgements}. Work of KP (settling accretion theory and its applications) is supported by RSF grant 16-12-10519. JMT acknowledges the research grant ESP2014-53672-C3-3-P and LO 
acknowledges the DLR grant 50 OR 1508.

\end{document}